LA-UR-08-06783

Title: Commissioning the DARHT-II Accelerator

Author(s): Carl Ekdahl, HX-6

Intended for: DARHT Technical Note 466

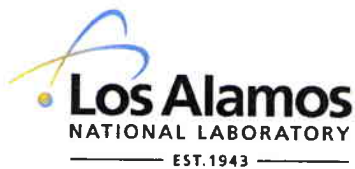

# Commissioning the DARHT-II accelerator

Carl Ekdahl, for the DARHT-II team

## I. INTRODUCTION

In its final configuration the 74-cell DARHT-II linear induction accelerator (LIA) produces a 2-kA electron beam. The beam kinetic energy is greater than 17 MeV in a pulse with a flat top greater than 1.6 µs. Up to four short pulses are cleaved out of this long pulse by the kicker and transported to the final focus where they are converted to bremsstrahlung for flash radiography of hydrodynamic experiments. We began commissioning the full-energy LIA on June 14, 2007 with shot 5618. On January 30, 2008, the beam stop was opened and the first full-energy beam injected into the kicker (shot 7471). This report summarizes the intervening work resulting in a beam acceptable for producing four radiographic pulses.

## II. BEAM DIAGNOSTICS

Since the first day, DARHT-II has been heavily instrumented with beam and accelerator diagnostics [1]. Some of these are non-invasive, and are used on every shot for monitoring injector/accelerator operation, and for adjusting the tune to give highest radiographic performance. Other diagnostics, such as beam profile imaging, are invasive and only used occasionally. Table I lists the accelerator and beam diagnostics. Figure 1. shows the locations of the beam position monitors (BPMs) on the full-energy accelerator. Figure 2. shows the locations of the diagnostics downstream of the accelerator exit, which were used to measure the parameters of the full-energy beam. Short descriptions of these diagnostics follow Figure 2.

Table I. Beam and Accelerator Diagnostics

| Parameter | Diagnostic | Detector | Number Installed | Data Channels |
|---|---|---|---|---|
| Diode Voltage | E-Dot | Flat Plate | 1 | 1 |
| Cell Accelerating Potential | CVM | Resistive Divider | 74 | 74 |
| Beam Current, Centroid Position and Motion | BPM | 4 or 8 Balanced B-dot loops | 20 | 240 |
| Beam Ellipticity | BPM | 8 Balanced B-dot loops | 2 | 64 |
| Beam Energy | Magnetic Spectrometer | Fiber coupled Streak Camera | 1 | |
| Beam Energy | Total Accelerating Potential | Diode E-dot + 74 CVMs | | |
| Beam Current Profile | Anamorphic streak | SiO2 plate | 1 | 4 |
| | | | | |
| | | | | |

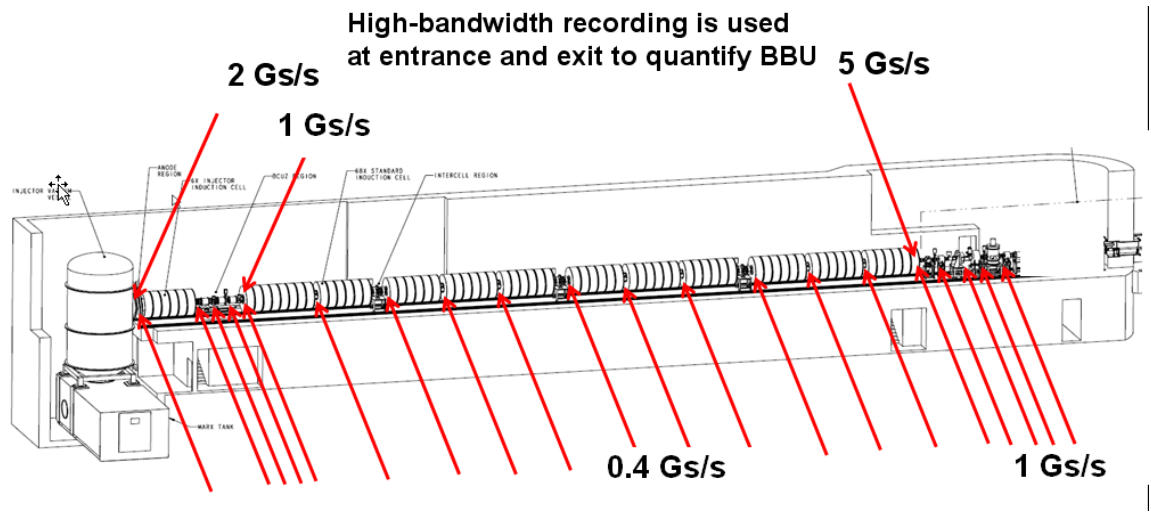

Figure 1. Location of beam position monitors for measuring beam current, beam centroid position and motion in the full-energy DARHT-II accelerator.

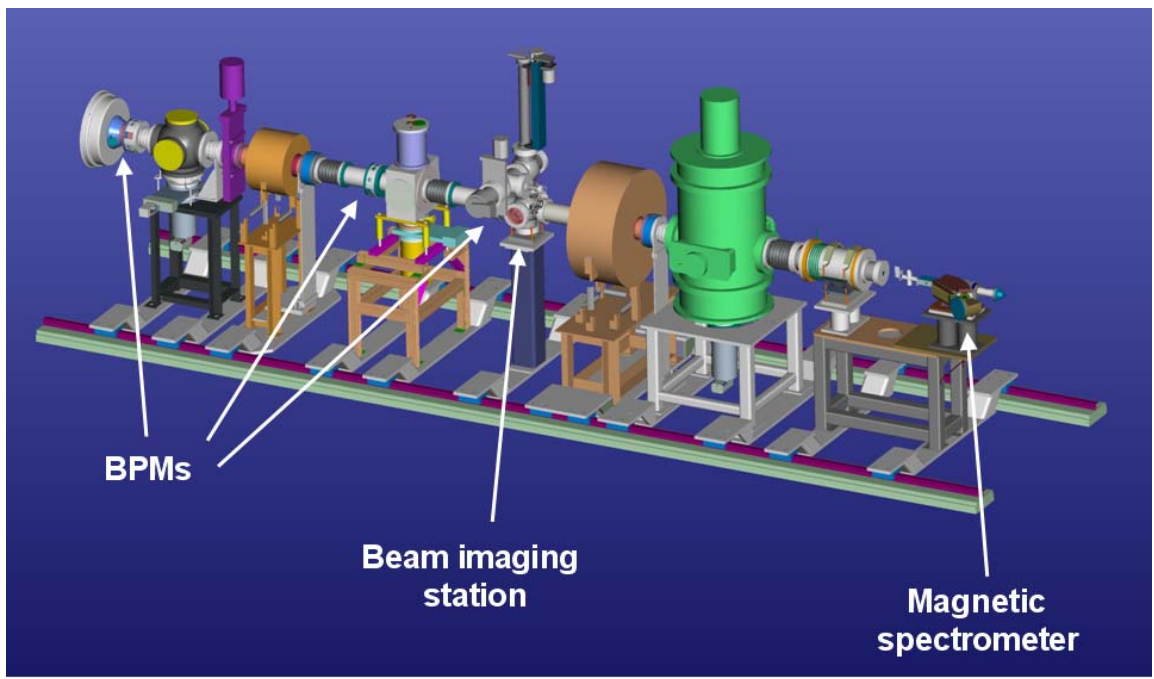

Figure 2. Diagnostics for measuring the DARHT-II full-energy beam parameters.

A. *Diode Voltage Monitors*

The A-K gap potential was measured with a capacitive divider. This monitor was a 19-cm diameter plate positioned flush with the vacuum tank wall. It faced the high-voltage dome containing the diode cathode structure. The location of the monitor was toward the rear of the dome, in a position that shielded it from electron emission and direct cathode light. The monitor was recorded directly, giving a signal proportional to dV/dt, which was integrated with software. The nominal calibration used for these measurements was that established by PSI.

### B. Acceleration-cell Diagnostics

In addition to diagnostics that monitor the injector Marx generator performance, each of the 74 induction-acceleration cells has a resistive divider to measure the voltage pulse delivered by the pulse-forming network (PFN). These cell voltage monitors (CVM) were summed and added to the diode voltage to estimate the accelerated beam energy.

### C. Magnetic Spectrometer

A 45° sector magnet powered by a highly regulated supply is used to measure the energy of beam electrons. The position of a collimated, 1-mm beamlet in the spectrometer's momentum dispersion plane was registered with an NSTec streak camera using light from a fast scintillator, which was focused onto a coherent fiber bundle that relayed it to a remotely located streak camera. This spectrometer system was calibrated with a negative-ion source at accurately measured ion energies (NIST traceable). Based on these calibrations, the beam measurements are accurate to within +/- 0.5% for electron kinetic energies between 3.0 MeV and 20.0 MeV. Measurements of beam energy using the spectrometer were then used to cross-calibrate the energy obtained by summing CVMs and the diode voltage. After removal of the spectrometer, the cell and diode voltage sum was used as a non-invasive monitor of the accelerated beam energy.

### D. Beam Position Monitors

Each beam position monitor (BPM) consisted of an array of four (or eight) $B_\theta$-field detectors spaced 90° apart (or 45° apart). Each detector was a balanced, shielded-loop design with a Moebius crossover to minimize common mode signals arising from ground loops, radiation driven Compton currents, direct beam spill pickup, electric field pickup, and/or other interfering noise, EMP, or backgrounds. Details of construction and analysis of data can be found in Ref. [1].

Calibration of the current and position sensitivity of each BPM was accomplished in a "coaxial" test stand with an inner conductor that could be accurately offset from center. These calibrations enabled us to make measurements of beam current with less than +/- 2% uncertainty, and position to better than +/- 0.5% of the beam-tube radius. The eight-detector BPMs were also used to measure beam ellipticity with an approximate uncertainty of 10-20% in the major-to-minor axis ratio; the somewhat larger uncertainty due to the need for estimates of the beam size from envelope code results.

### E. Beam Imaging

The beam current profile was imaged 3.2 meters after exiting the accelerator. The beam generated Cerenkov light from thin (sub mm) $SiO2$ targets oriented at 45° to the beam axis. The targets were frosted on the back surface to scatter Cerenkov radiation into the line of sight orthogonal to the beamline where it was imaged with a Bechtel-Nevada streak camera system. Target insertion, rotation, and manipulation was accomplished remotely from the control room.

The Bechtel-Nevada streak camera system used unique, anamorphic optics that compressed the light in one dimension into a line focus on a coherent, linear fiber-optic bundle. The other end of the coherent bundle was cemented to the remotely located camera. Four such lens/coupled fiber arrangements provided simultaneous projections in the horizontal (X), vertical (Y), + 45°, and -45° orientations. These were recorded on two 1024x1024 CCD readout streak cameras with two fiber bundles per camera. The anamorphic optical system simplified alignment, eliminated ambiguity resulting from beam motion, and eased analysis [2]. Anamorphic compression of the entire field of view in each direction simplified the calculation of moments of the beam distribution in that direction. The compression amounts to an optical integration in the direction orthogonal to the line image, and all that remains to be done is to compute the required moment along the direction of the line at each time. Moreover, maximum-entropy tomographic reconstructions of the streak data from this system provided motion pictures of the evolution of the beam current profile [2].

### F. Miscellaneous

There were a number of diagnostics used in addition to the major beam and accelerator diagnostics listed above. Examples are "open shutter" CCD cameras to localize light from possible column arcing, a blue-filtered PMT observing light from the diode region, a CCD imaging the hot cathode through and array of selectable filters to observe possible temperature variations across the surface, and an optical pyrometer to measure the cathode temperature in a single spot.

### III. Magnetic Transport

#### A. Overview

The tunes for the magnetic transport were designed using four separate beam dynamic codes. Two of these were codes that solve the beam-envelope equations describing the beam envelope [3] through the accelerator. The beam-envelope differential equations were solved for beam parameters and initial conditions at a beam launch point downstream of the diode: Electron kinetic energy (KE), beam current (Ibeam), equivalent-envelope radius (r0), envelope convergence (r0p), and normalized emittance (en). The equivalent envelope radius is 1.414 times the rms radius for a non-uniform beam, and the normalized emittance is 4 times the rms emittance.

Most of this work was done with the XTR beam-envelope code [4], which produces the beam envelope through the accelerator for a single initial energy. XTR can also solve the differential equations describing the beam centroid motion, and it was used to predict the effects of initial beam offset and steering with dipoles.

LAMDA [5] is also an envelope/centroid-motion code that solves the equations for successive disks of a time-segmented beam to obtain time resolved envelope/centroid-motion results. The versions of XTR and LAMDA that we used during commissioning incorporated exactly the same physics models, and their results were identical for identical initial conditions, magnetic tunes, and pipe geometries. We used LAMDA to predict the transported pulse waveform for different tunes of the BCUZ, and the time-resolved head/tail motion to be expected from initial beam offset.

Both of these codes are essentially differential equation solvers, and so they must be supplied with accurate initial conditions for trustworthy results. So far, only one of the parameters, beam current, has been directly measured with well-calibrated diagnostics. The initial beam energy can be inferred from the diode voltage measurements, but the calibration of this diagnostic is not as sound. As for the initial conditions (r0, r0p), and initial emittance (en), we relied on simulations of the diode using the TRAK gun-design ray-tracing code [6] and the LSP particle-in-cell (PIC) code [7]. We built tables of the initial conditions at the envelope-code initial position (100 cm, in LBNL coordinate system) for 100-kV steps of the diode voltage, assuming space-charge limited emission from the dispenser cathode. This is the same procedure used for all previous DARHT-II experiments, but since the diode was modified to a new, high-perveance geometry we had to repeat all of the steps in the process. The initial position (100 cm) was chosen to be far enough downstream of the anode entrance that fringe fields from the applied diode voltage were significantly less than the beam space-charge fields [8]. This ensured that the beam dynamics at the position of handoff to the envelope codes were uninfluenced by the applied diode voltage, which is not modeled in those codes.

#### B. Diode Simulations and Measurements

The diode simulations require the magnetic fields in the diode as input. These fields are also required input for the envelope codes. To tune the injector for acceptable transport, they were adjusted to ensure that the maximum extent of the beam envelope was less than 80% of the pipe radius for all energies in the beam head, even with the cells un-energized. (This "guarantees" lossless transport through the injector, even in the worst case.) Therefore, the process of tuning the injector for acceptable transport by adjusting the solenoidal focusing fields required iterating back and forth between the diode modeling and the envelope codes to obtain self-consistent initial conditions. The initial parameter tables that resulted were then used for investigating off energy beam transport and centroid motion with XTR. They were also used for LAMDA predictions of current pulse waveforms interpolating to experimental diode voltage waveforms.

The following few illustrations show the space-charge limited beam predicted by the simulations of the high-perveance diode geometry, followed by measurements of diode parameters for this geometry

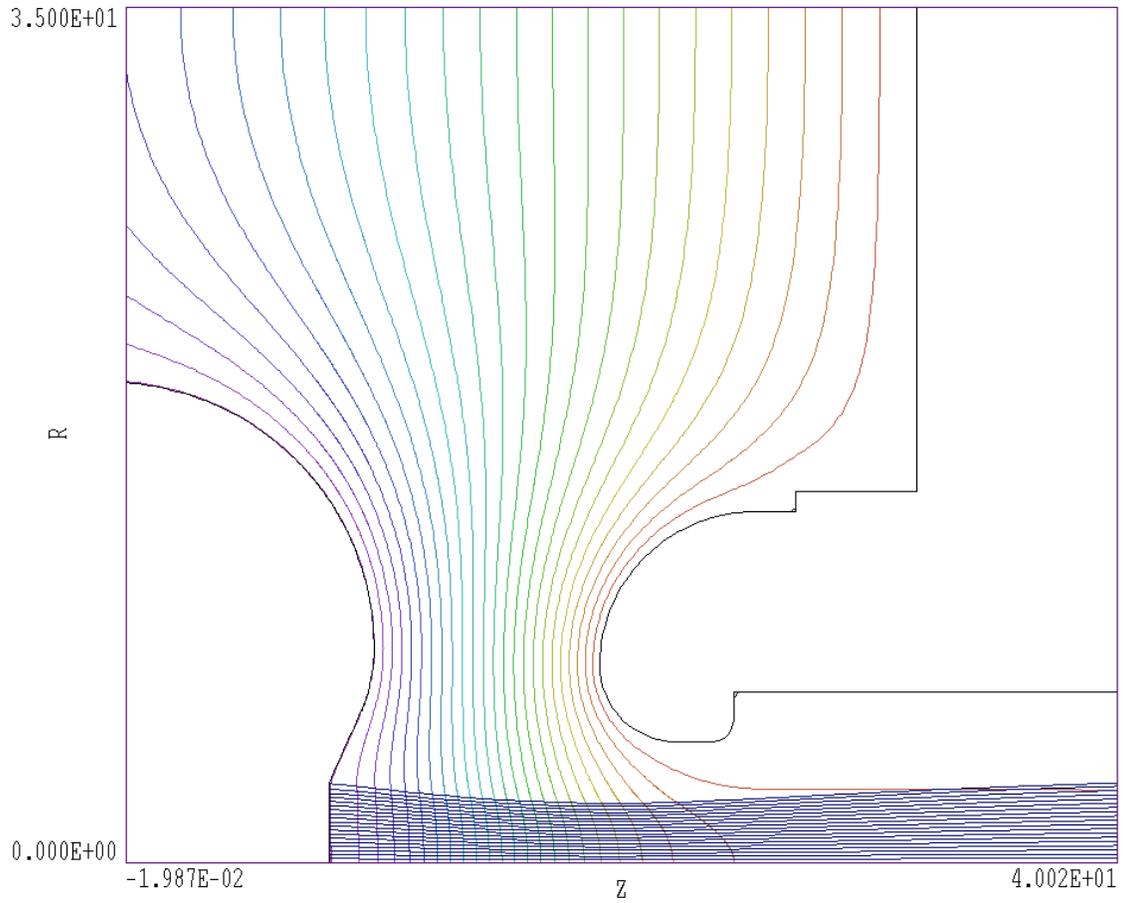

Figure 3. TRAK simulation of the high-perveance diode. Dimensions are in inches. The right hand boundary approximately corresponds to the starting point of the envelope codes. The equipotentials displayed are the sum of the applied diode plus beam space-charge.

Figure 4. LSP simulation of the high-perveance diode.

Figure 5 shows the diode AK voltage measured by the dV/dt voltage monitor on shot 7426, near the end of the accelerator commissioning. The calibration factor applied to the signal for these data was adjusted down by 1.7% from the nominal PSI calibration in order to make the flat-top voltage agree with that required in the simulations to produce a space-charge limited current equal to the measured current (Figure 6).

The diode perveance (I/V^3/2) calculated from these measurements is shown in Figure 7. Also shown in this figure is the space-charge limited perveance from the TRAK/LSP simulations. The close agreement between measurement and simulation over the flat top has been forced by the aforementioned adjustment of the diode voltage-monitor calibration factor. The range of diode voltage for which the perveance is plotted is -0.5 MV to -2.53 MV. The close agreement between measurement and simulation over this range of voltages during the pulse risetime is strong evidence for space-charge limited emission from the cathode.

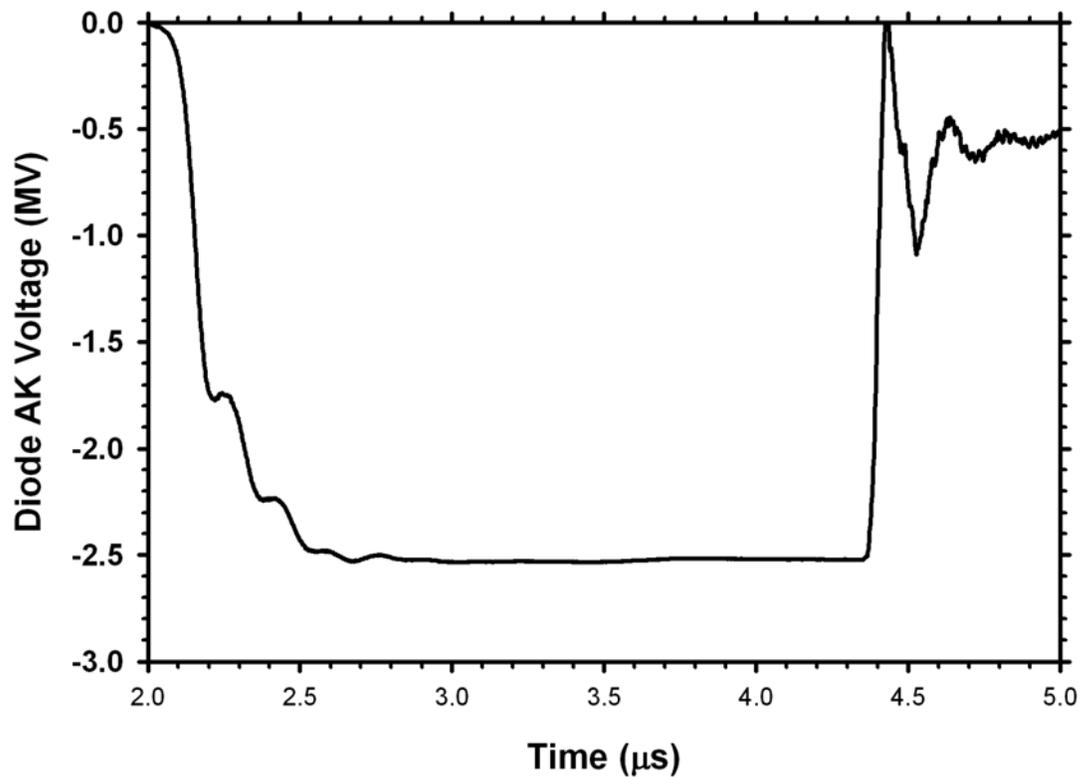

Figure 5. Diode Voltage measured for shot 7426. Calibration adjusted to agree with space-charge limited current shown in Figure 4.

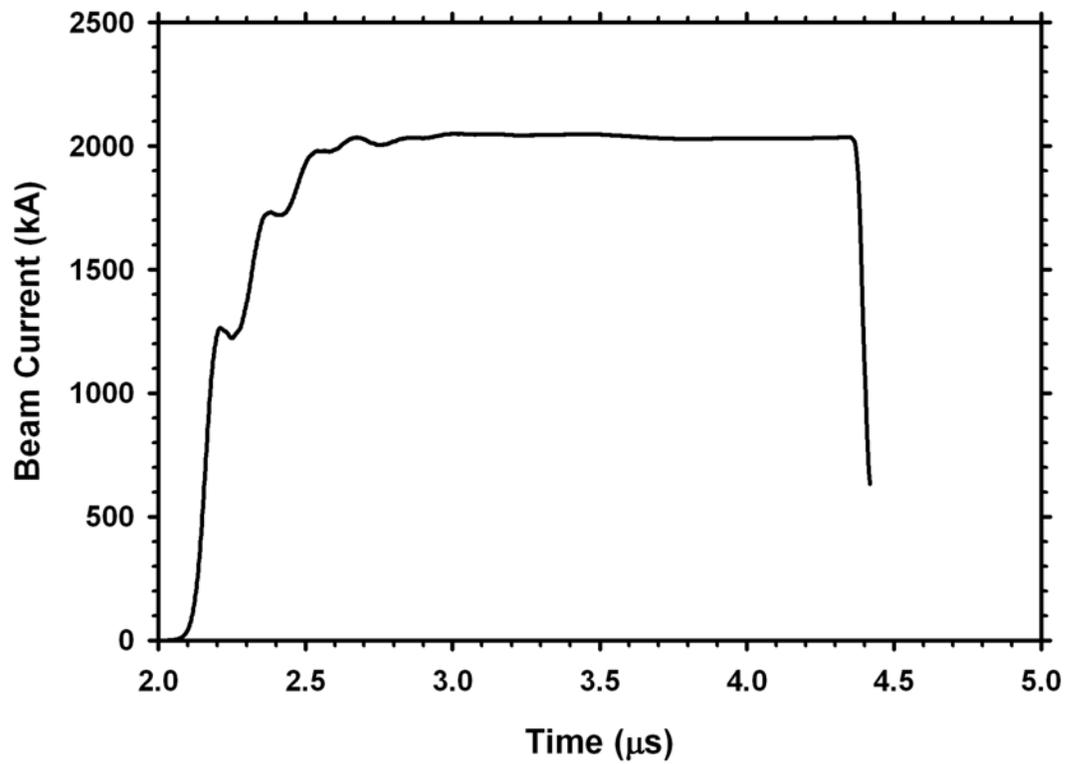

Figure 6. Current extracted from diode on shot 7426.

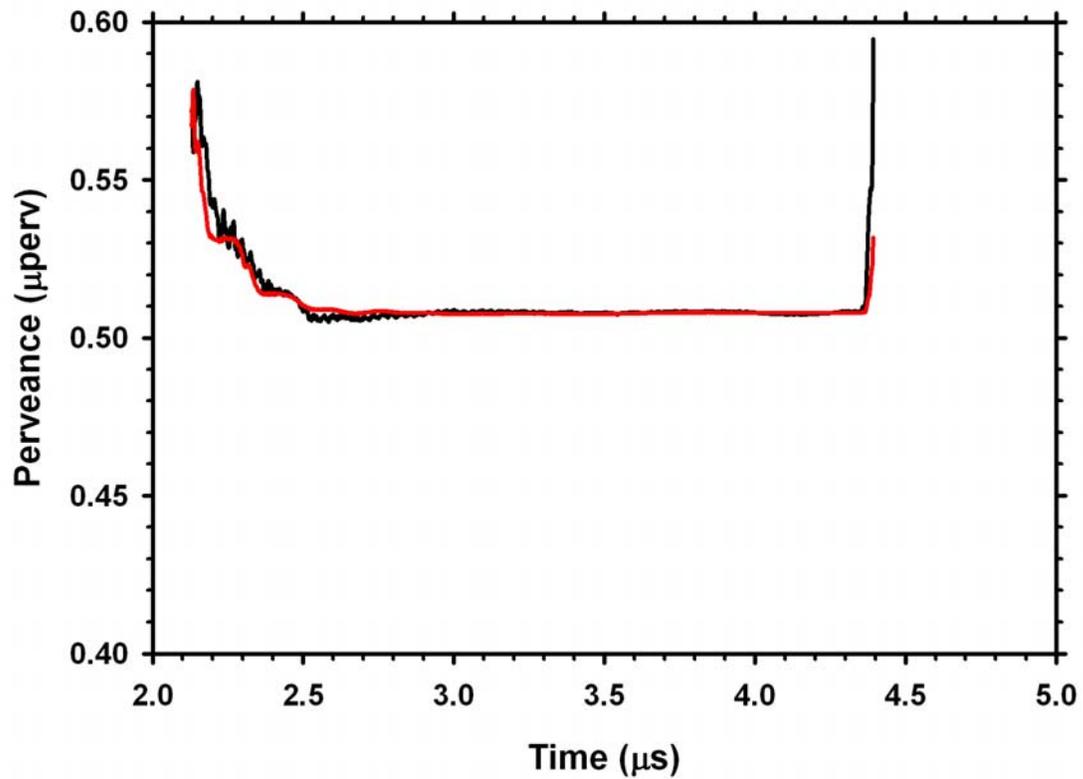

Figure 7. Diode perveance from TRAK/LSP simulations (solid red) compared with experimental data for shot 7426 (solid black).

*C. Cleanup Tune Simulations and Measurements*

In daily operations we always began with a relaxed tune that scraped of almost no beam current in the BCUZ, and a pulse shortened by the crowbar to give only ~200-ns of "flattop." After establishing adequate diode/accelerator/diagnostics performance with these settings we changed parameters to develop a tune that would produce an accelerated beam meeting the minimal requirements for safe injection into the kicker, downstream transport, and final focus. Initially, we sought a tune with the off-energy head and tail scraped off and lost in the BCUZ.

The magnetic tune designed to scrape off the off-energy beam head in the BCUZ is shown in Figure 8, and Figure 9 shows a comparison between simulation and BPM measurements for this tune.

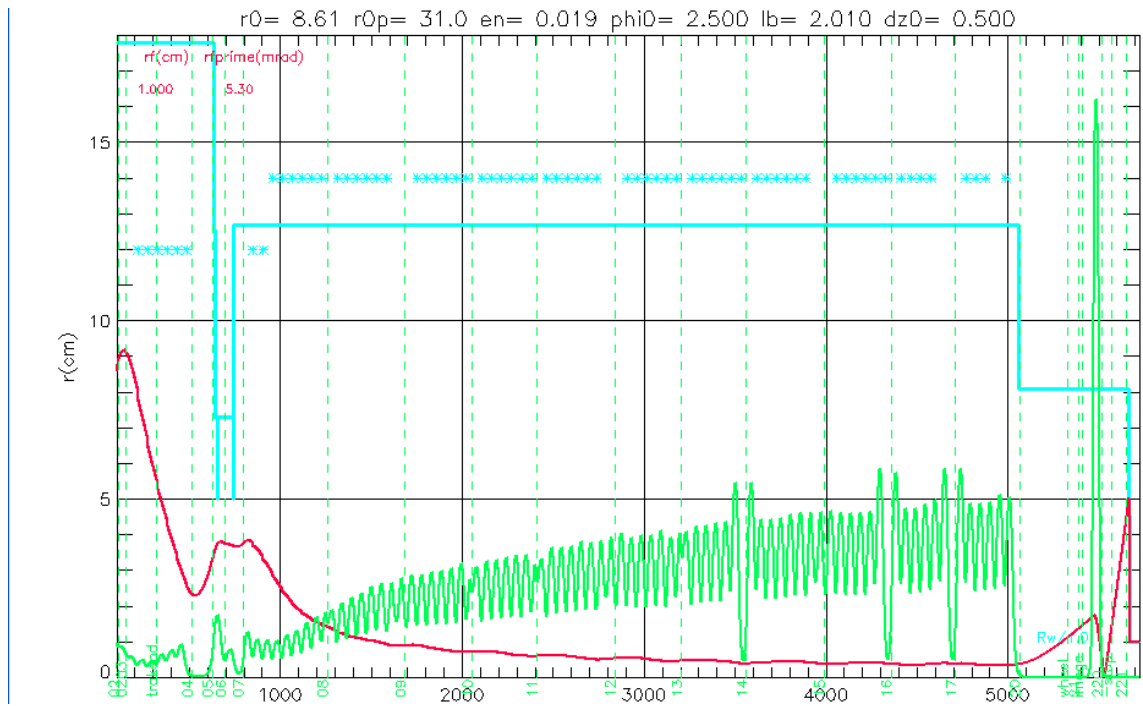

Figure 8. XTR simulation of the cleanup tune. The z-zxis units are cm. The solid red curve is the beam envelope radius, the solid green curve is proportional to the solenoidal focusing field on axis. The vertical dashed green lines indicate the positions of the BPMs located at the entrance to each cell block. The solid blue line is the radius of the beam pipe, also shows the BCUZ apertures. The blue asterisks are proportional to the accelerating potential of each cell.

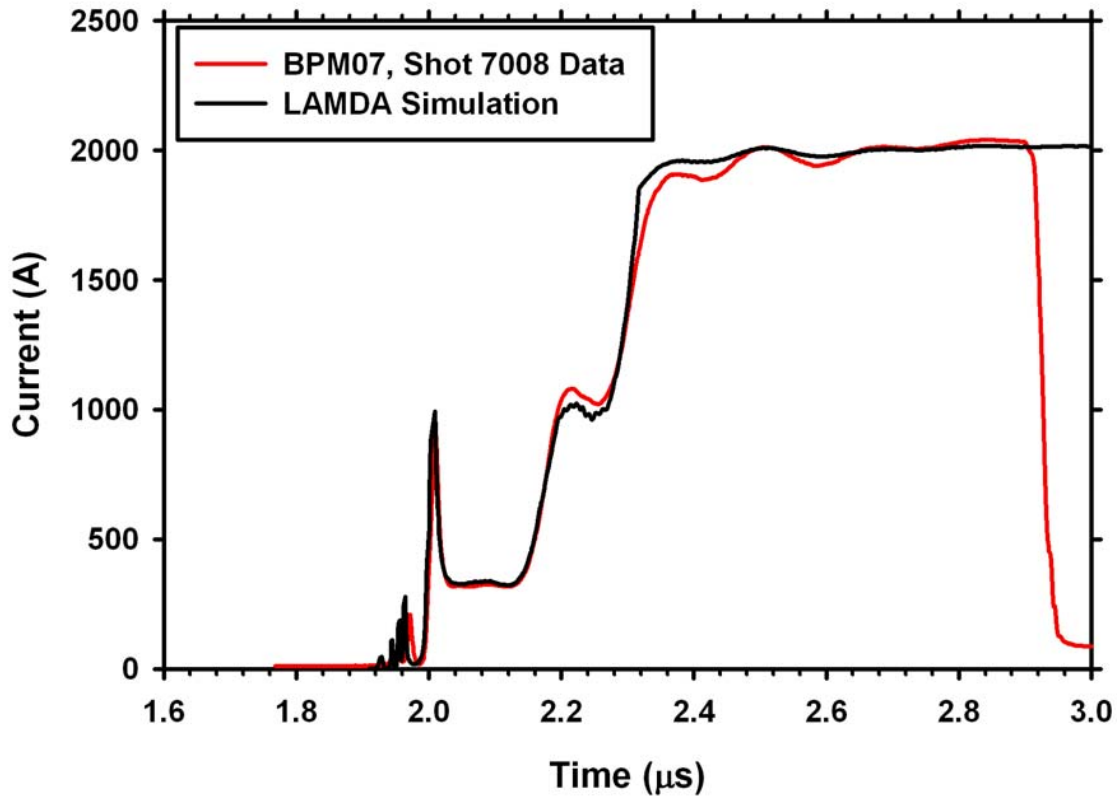

Figure 9. Current transmitted through the BCUZ for the cleanup tune. The shorter pulse (shown in red) is data from shot 7008 from the first BPM beyond the BCUZ. The longer pulse (shown in black) is the result of the LAMDA simulation.

Figure 9 shows the measured current transmitted through the BCUZ compared with a LAMDA simulation. This simulation used TRAK/LSP initial conditions, the calculated magnetic focusing fields of the six injector cells and the BCUZ, the injector cell accelerations, and the BCUZ apertures. The agreement between the data and the simulation validates the procedures that we use for obtaining self consistent initial conditions from TRAK and LSP and applying them to the envelope codes.

In Figure 10 we show the current transported through the accelerator by this cleanup tune for a the same shot as Figure 9, with a short crowbar delay. (BPM data following the fast crowbar was unreliable due to the limited dynamic range of the data recording instrumentation, so is not displayed in these plots.) In Figure 9 the vertical red markers delineate the 1.6-μs flattop region later used for commissioning the four-pulse target. Attempts to stretch the cleanup tune to this full width were largely unsuccessful because of an interaction between the aggressive beam scraping and the diode, which led to unreliable late-time diode performance. In the limited time available, we tried other cleanup tunes and tactics to no avail, and finally elected to use the relaxed, start-up tune for commissioning the kicker, downstream transport, and four pulse target.

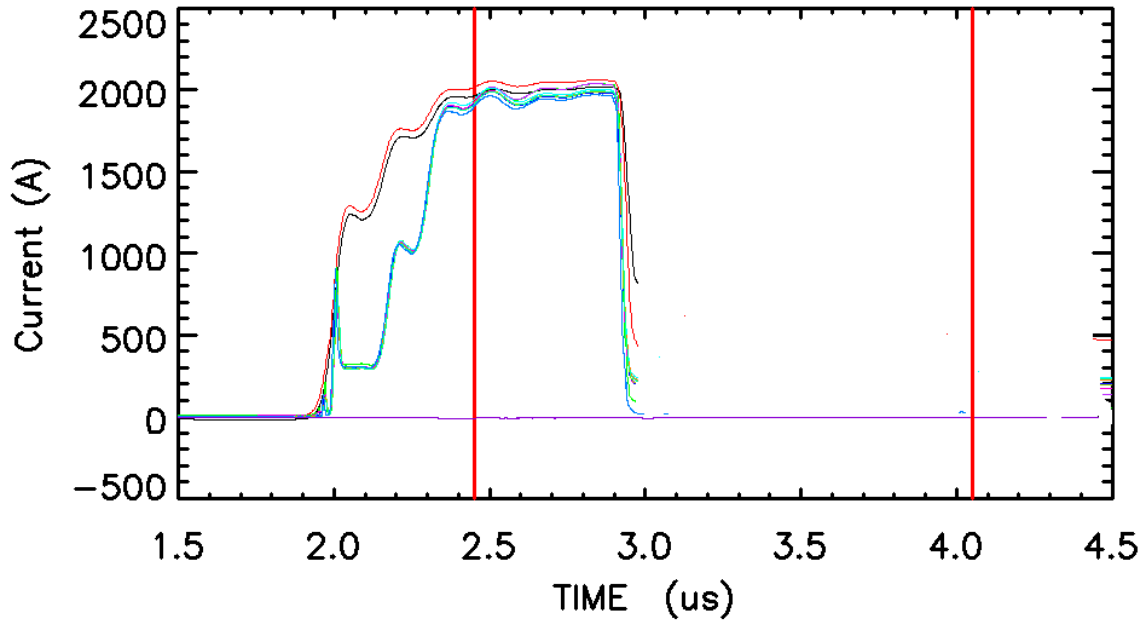

Figure 10. All BPMs cleanup

*D. Relaxed Tune Simulations and Measurements*

The accelerator tune used for commissioning the kicker, downstream transport, and four-pulse target was designed to transport almost all of the beam through the BCUZ without scraping off much of the off-energy head. The XTR simulation of this "relaxed" tune is shown in Figure 11. Once the beam passes through the BCUZ it is quickly focused to the small matched envelope dictated by rapidly increasing magnetic field to suppress the BBU. After exiting the accelerator, the beam expansion in the drift region is insufficient to increase the size enough to prevent overheating the beam-stop during long-pulse operations. Since overheating (> 300-400 C) can lead to ion focusing and damage the beam stop, the S-2 solenoid was incorporated in the design blow it up through over focusing, as shown in the figure.

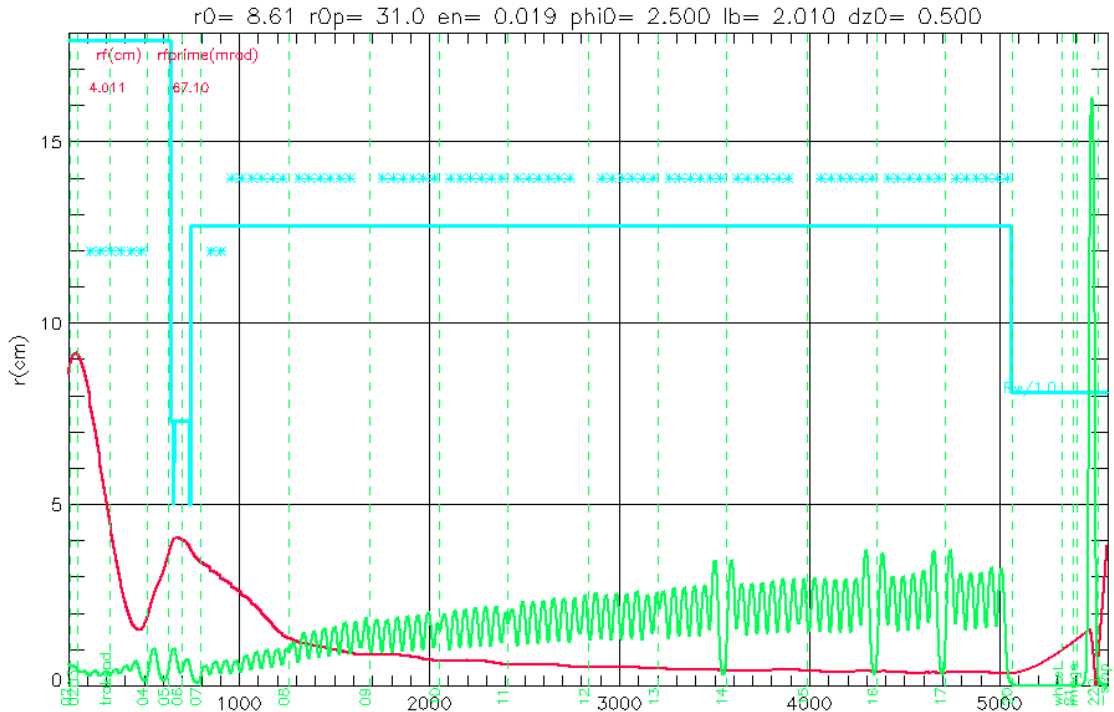

Figure 11. XTR simulation of the relaxed tune.

During commissioning we lost some of the accelerating cells to flashovers and vacuum breakdowns. Turning these cells off on subsequent shots resulted in loss of full acceleration. Since the difference between PFN pulse voltage and acceleration potential is due to beam loading, turning a cell off reduces the acceleration by the full pulse voltage. That is, with the cell off, the beam loading (~13 keV/kA) results in deceleration. Cells turned off was accounted for in the envelope-code tuning simulations. At the conclusion of commissioning, we had four cells turned off (26, 61, 62, and 69). As shown in Figure 12, turning these four cells off made very little difference in the envelope compared with Fig. 11, for which all cells were turned on. The most difference was from cell 26, the last cell in cell-block 3, where the beam energy is still low enough that the cell loss kicks off a mild betatron oscillation of the envelope.

With the relaxed tune, there was no loss of beam current through the accelerator during the time that the accelerating cells were energized. The transport of the beam through the Figure 13 is an overlay of beam current measurements at the entrance and exit of the injector cells and through the accelerator. The crowbar was timed to coincide with the end of the accelerating cell pulse. During the 1.6-μs flattop region later used to commission the four-pulse target (shown by the red cursors) the average current was 2112 A with a 1.2% standard deviation. (The standard deviation of the current at each BPM during the flattop was less that 1%.)

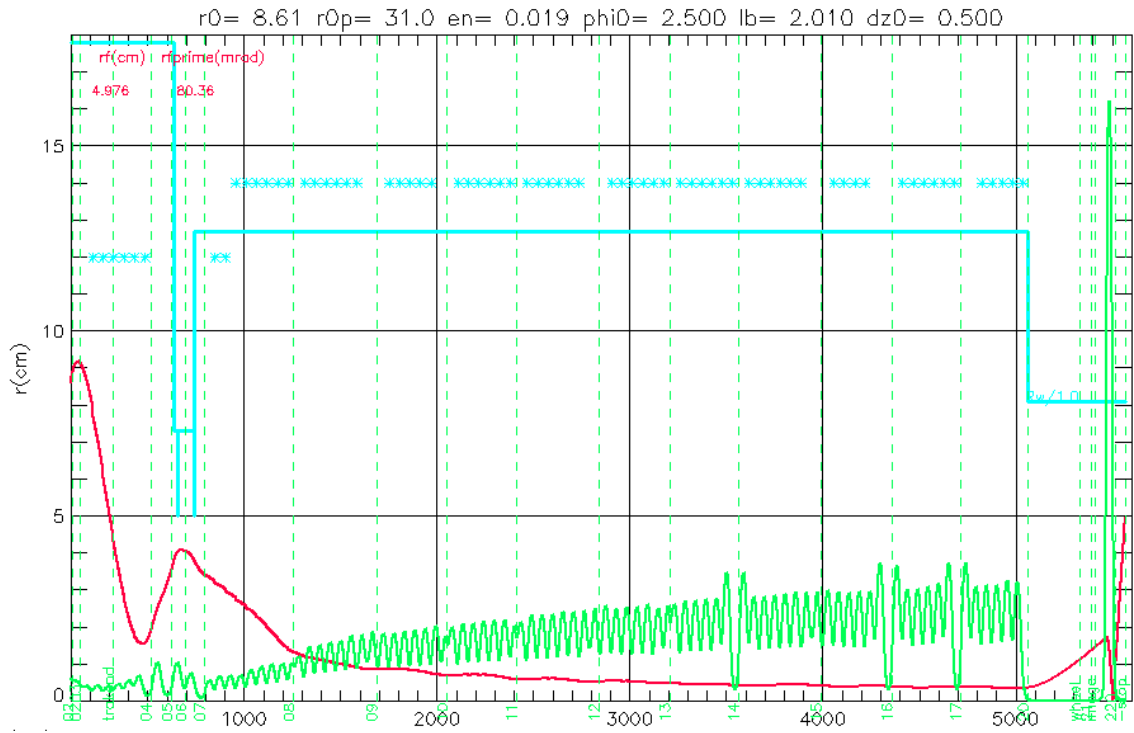

Figure 12. XTR simulation of the relaxed tune with cells 26, 61, 62, and 69 turned off (decelerating potential of 26-keV per cell).

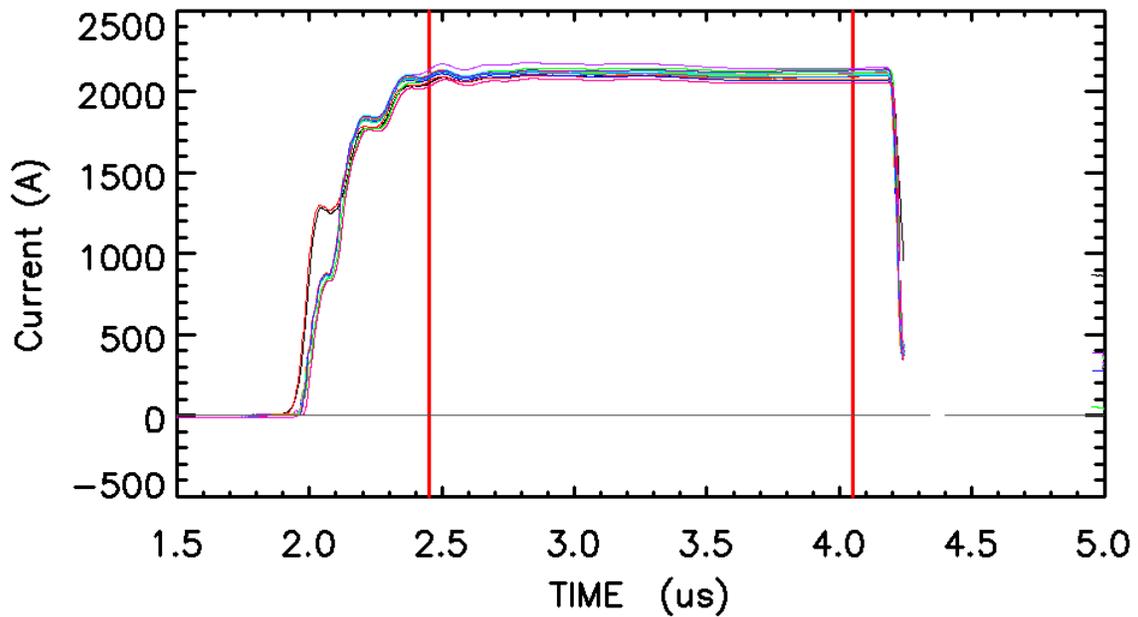

Figure 13. Overlay of beam current measurements in injector and accelerator for the relaxed tune for a shot near the end of accelerator commissioning (shot 7426, 1/23/08). The red cursors indicate the 1.6 µs region later used for the four-pulse target commissioning.

13 of 26

In order to reduce the beam sweep at the accelerator exit the corrector/steering dipoles were used to more accurately center the beam at the BPMs located at the entrance to the cell blocks. Figure 14 shows the beam displacement from the axis averaged over the 1.6-μs window shown in Figure 13. Also shown is the rms variation of displacement, which is a quantitative measure of corkscrew and sweep. This shot (7169) was taken before the steering was applied. The improvement after applying steering is shown in Figure 15.

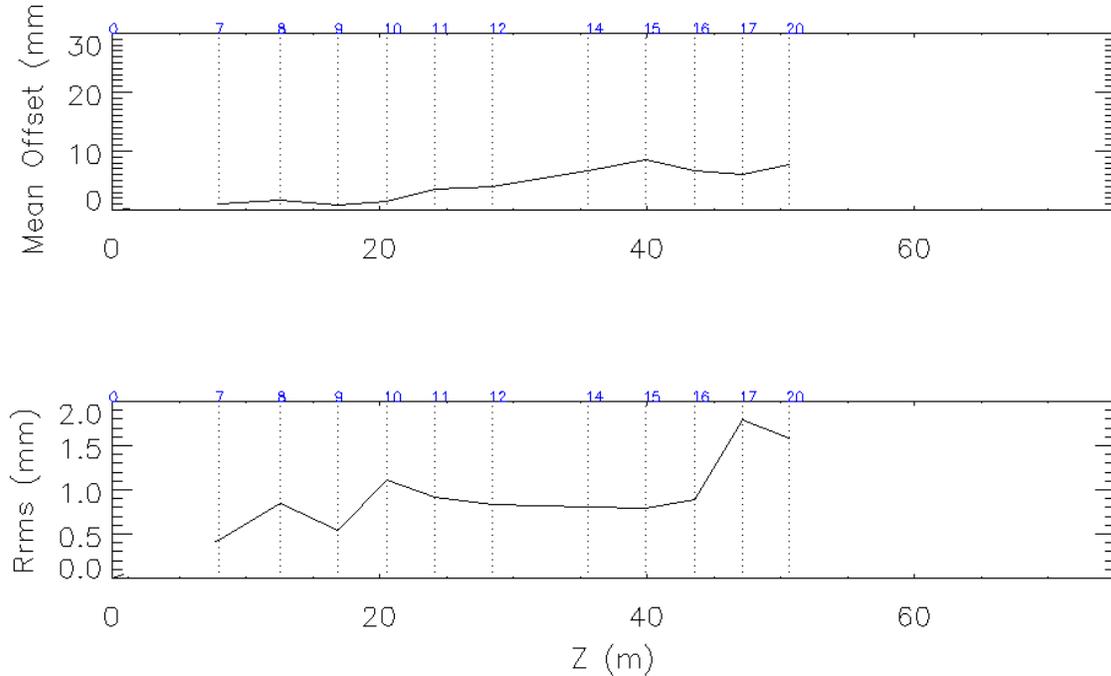

Figure 14. Top: Beam position throughout the accelerator on shot 7169 (no steering) averaged over the 1.6-μs window shown in Figure 13. Bottom:The rms variation of displacement, which is a quantitative measure of the extent of motion during the window.

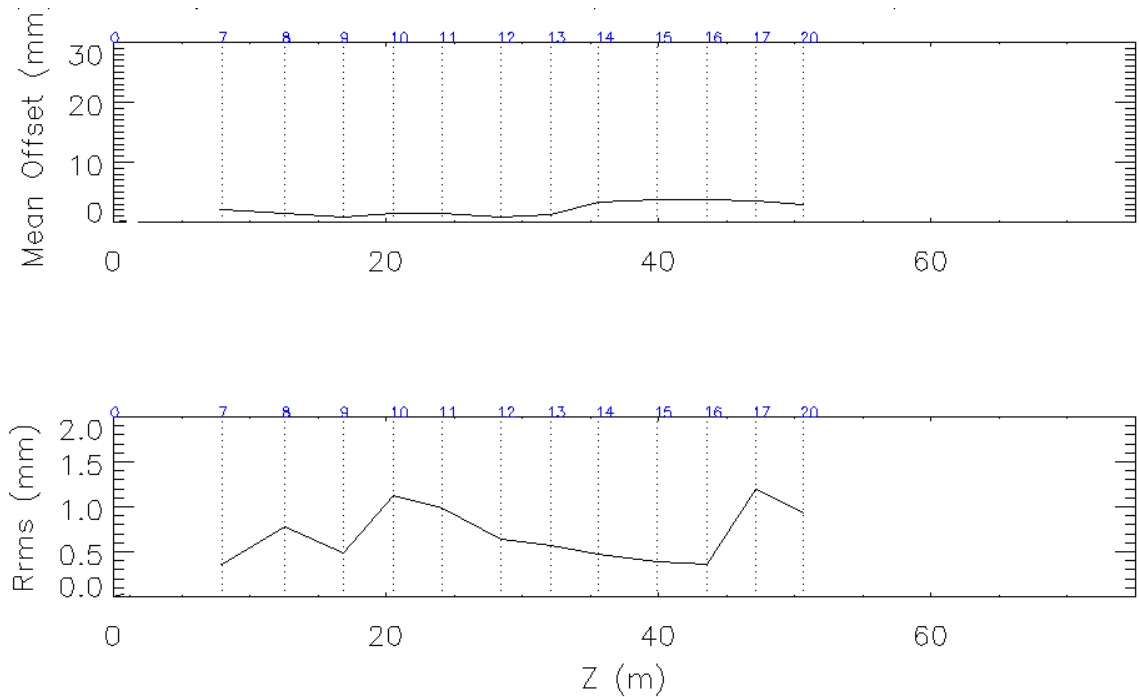

Figure 15. Top: Beam position throughout the accelerator on shot 7426 (steering applied) averaged over the 1.6-μs window shown in Figure 13. Bottom: The rms variation of displacement, which is a quantitative measure of the extent of motion during the window.

IV.  ACCELERATED BEAM PARAMETERS

*A.  Accelerated-Beam Parameters*

The accelerated beam parameters were measured after it left the accelerator with the diagnostics shown in Figure 2. The beam current is shown in Figure 16.

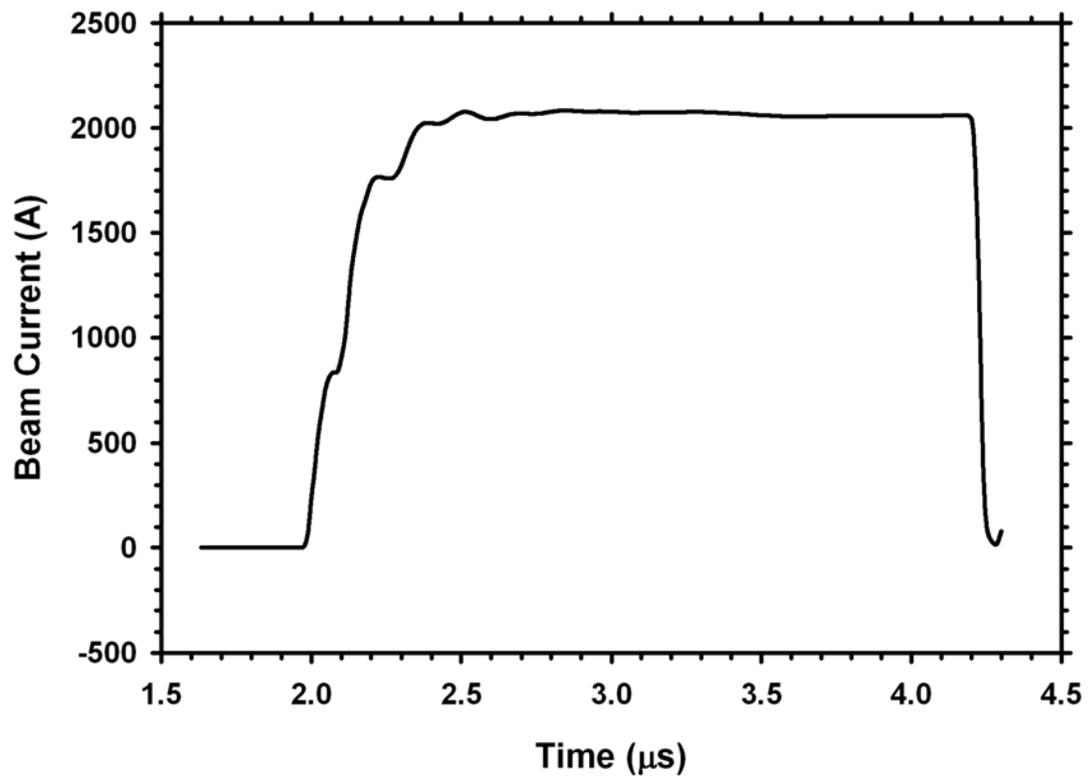

Figure 16. Beam current measured with BPM20 located just after the exit of the accelerator.

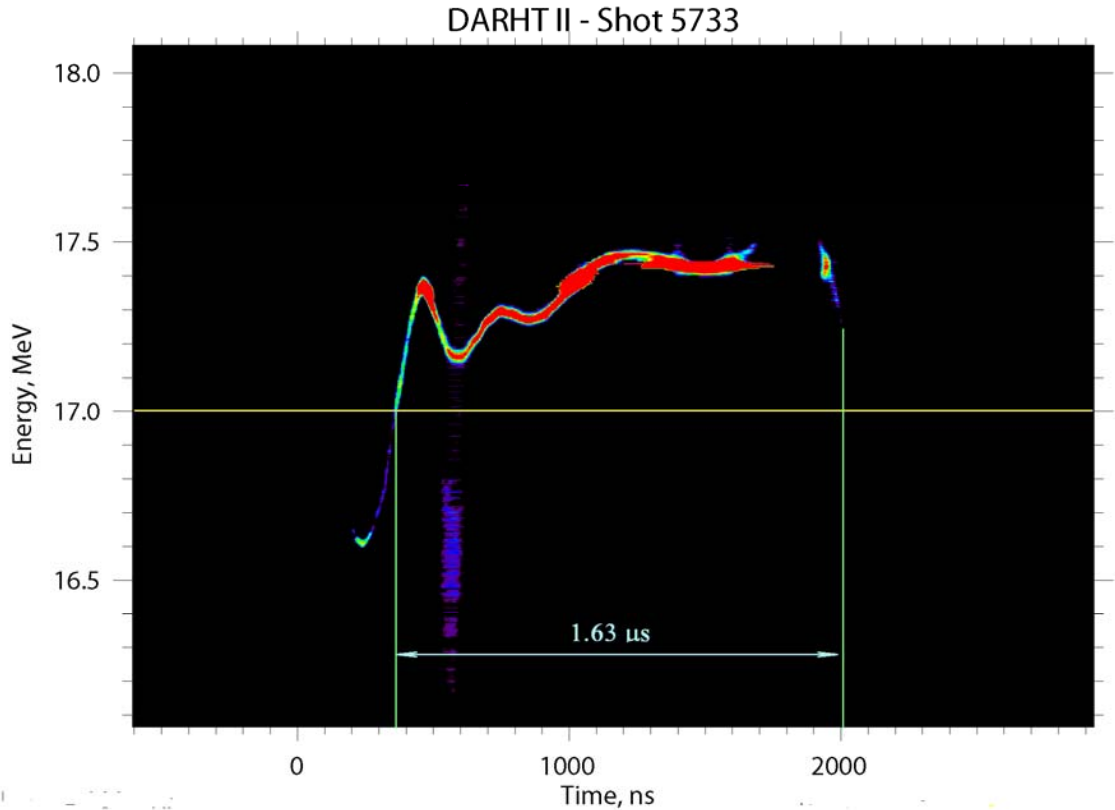

Figure 17. Magnetic spectrometer measurement of electron kinetic energy for shot 5733. On this shot five of the cells were turned off, which reduced the energy by ~1.25 MeV from what would be expected with all 74 cells.

Figure 17 shows the magnetic spectrometer measurement of electron kinetic energy for shot 5733. It can be clearly seen that the kinetic energy beam of the accelerated beam exceeds 17.0 MeV for more than 1.6 µs. On this shot five of the cells were turned off, which reduced the energy by ~1.25 MeV from what would be expected with all 74 cells.

A later measurement of the beam energy with a reduced cell charge voltage is shown in Figure 18. For this shot, six of the cells were turned off, which reduced the energy by ~1.5 MeV from what would be expected from all 74 cells at this reduced charge voltage.

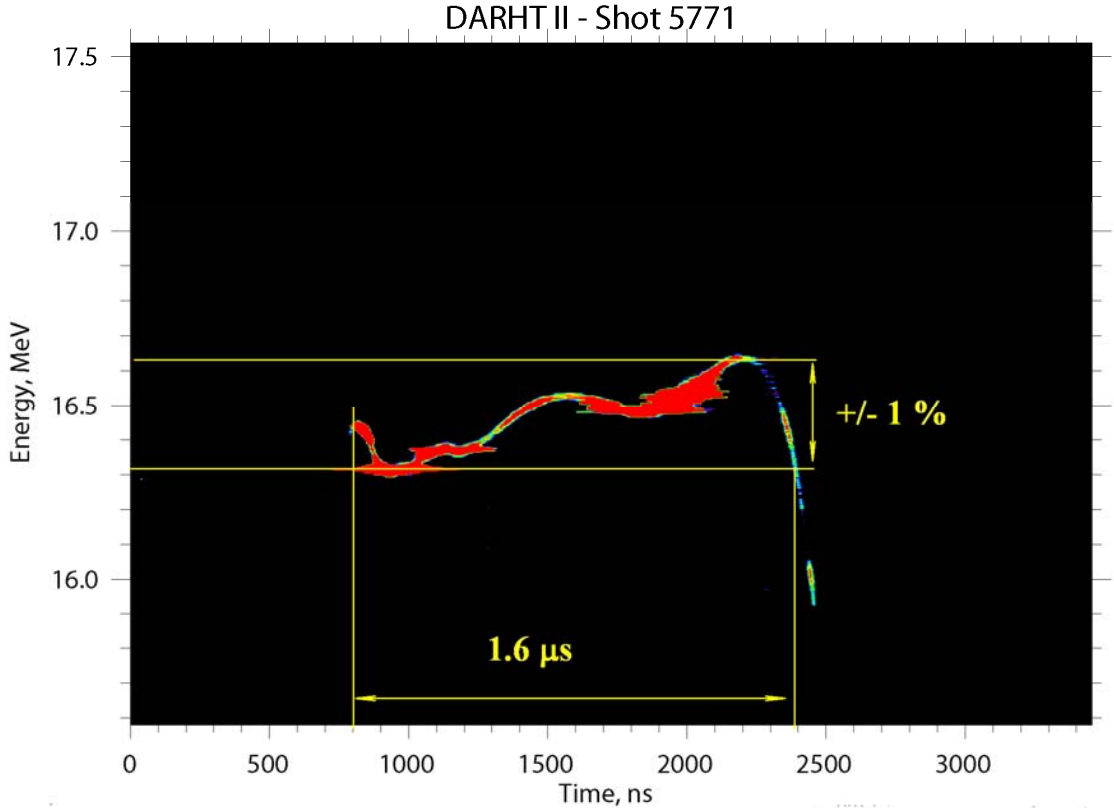

Figure 18. Magnetic spectrometer measurement of electron kinetic energy for shot 5771. On this six shot of the cells were turned off, which reduced the energy by ~1.5 MeV from what would be expected with all 74 cells.

### B. Beam Motion

The beam motion at the exit of the accelerator was dominated by an energy dependent sweep. The extent of the sweep is shown in Figure 19 for a shot early in the accelerator commissioning. The range of time for these plots is the 1.6-µs flattop region shown in Figure 13. The two leftmost plots show the range of beam motion (mm) at BPM20, just after the accelerator exit. The middle two plots show the range of motion (mm) at BPM21, which is 2.8-m downstream of BPM20. The rightmost plots show the angular sweep (mr) inferred from these two BPMs. The detailed variations in sweep (especially in y at BPM20 in x at BPM 21) at these two BPMS are well correlated with the variations in accelerating cell voltage waveforms, which is clear evidence for the energy dependence of this beam motion.

To reduce the sweep at the exit we steered the beam through the accelerator to more nearly center it at the entrance to each cell block. We also varied the steering in the last cell block to minimize the sweep (partial "tuning V"). The 1.6-µs sweep after these tuning modifications is shown in Figure 20 for shot 7426. This minimal effort expended to reduce the sweep resulted in a ~40% reduction in position sweep and ~30% reduction in angular sweep, which was deemed acceptable for commissioning the downstream and four pulse target.

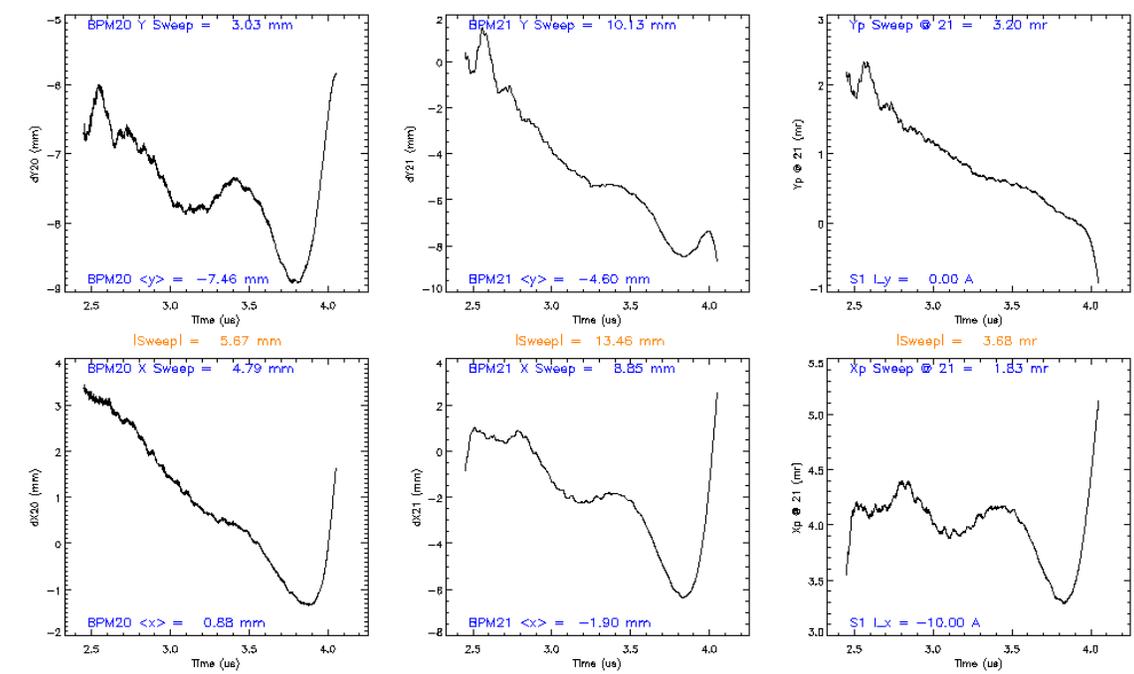

Figure 19. Beam sweep at accelerator exit for shot 7169. The time window corresponds to the 1.6-μs flattop region shown in Figure 13.

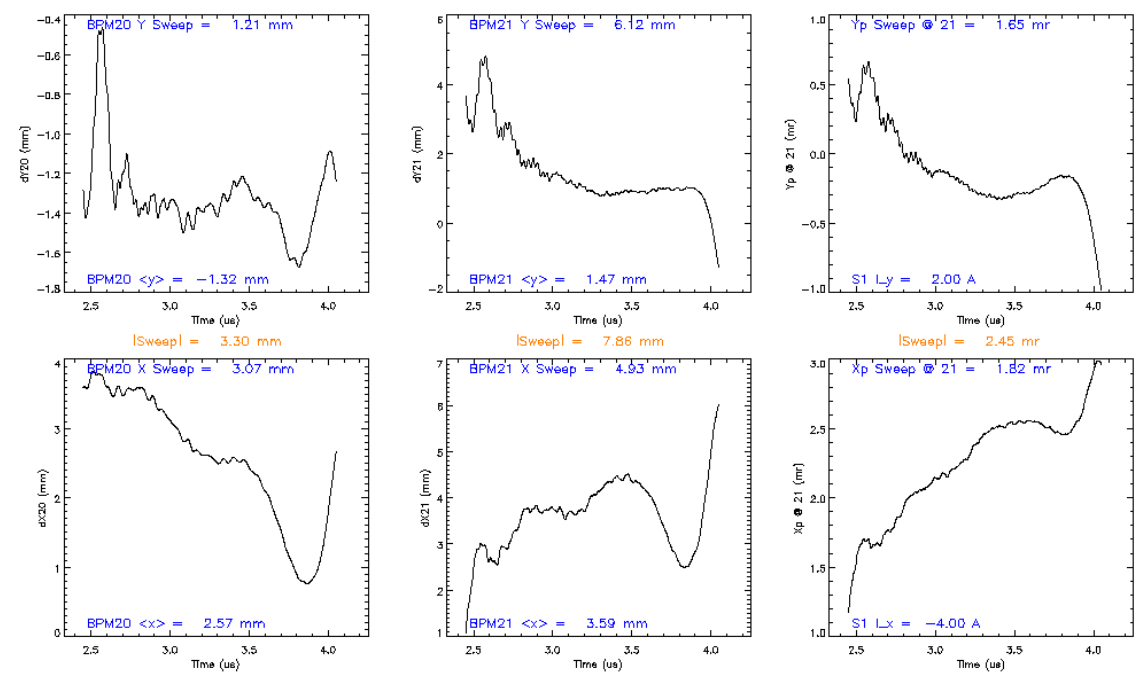

Figure 20. Beam sweep after reducing it by steering the beam through the accelerator. These are data for shot 7426. The time window corresponds to the 1.6-μs flattop region shown in Figure 13.

## C. Stability

As seen in Figure 20, the beam motion at the exit of the accelerator is dominated by a slow sweep imprinted with a low frequency (~ 1.5 MHz) oscillation characteristic of the accelerating cells (see Figures 17 and 18). In addition, there are lower amplitude perturbations with frequencies below that of the lowest BBU mode. The amplitude of this intermediate frequency motion is shown in Figure 21, for which the sweep and low frequency motion has been removed with a seventh degree polynomial fit. These intermediate frequency disturbances, with amplitudes less than ~500 microns, may be the result of ion hose caused by gas evolved in the BCUZ as the beam head is scraped. The frequency range of this motion (10 MHz to 20 MHz, see Figure 22) is approximately that predicted for ion-hose on a beam with the nominal DARHT-II envelope size predicted by XTR for this tune [9, 10].

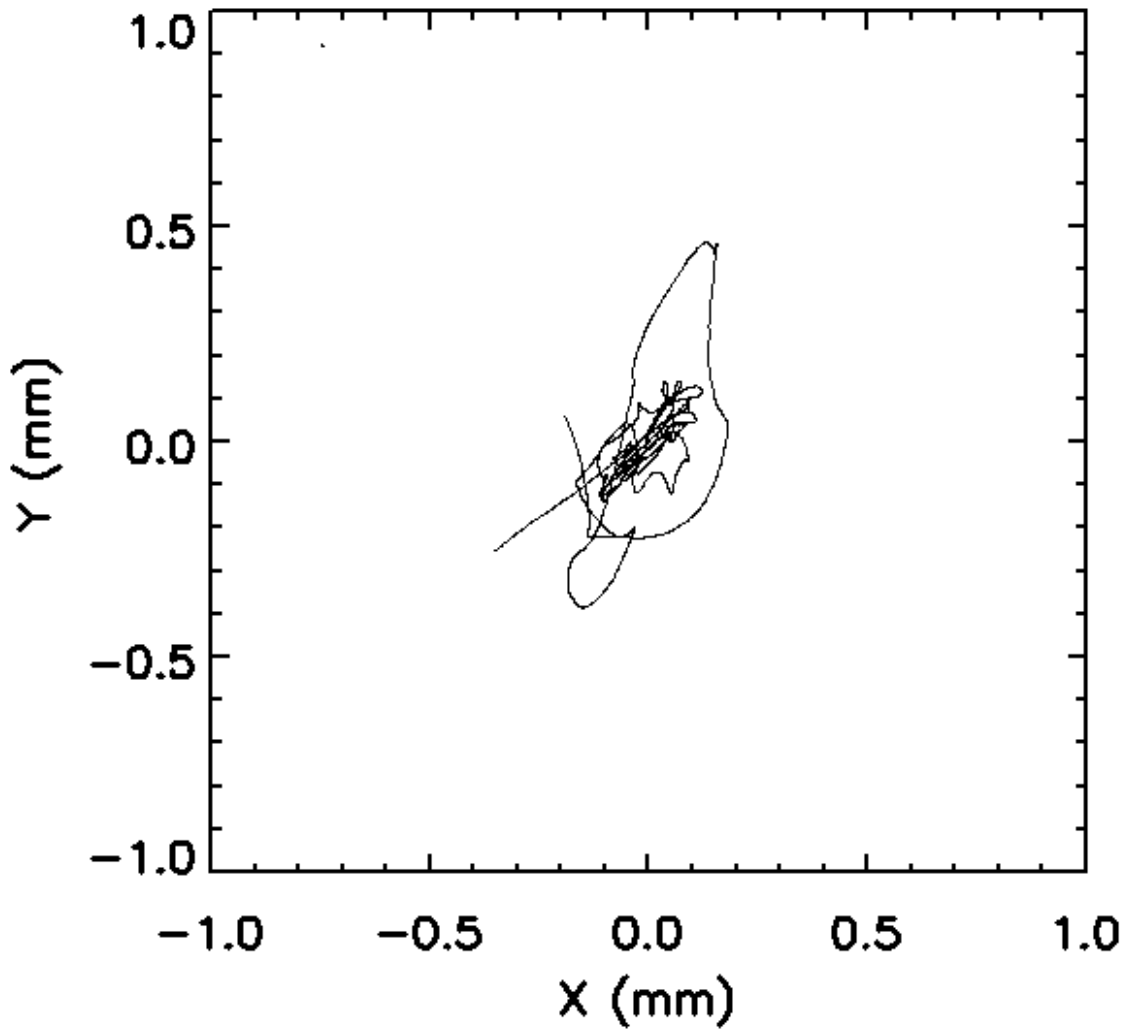

Figure 21. Intermediate-frequency corkscrew motion at the accelerator exit for shot 7426 during the 1.6-µs region shown in Figure 13. (The sweep and low frequency motion has been removed from the data with a seventh degree polynomial fit.)

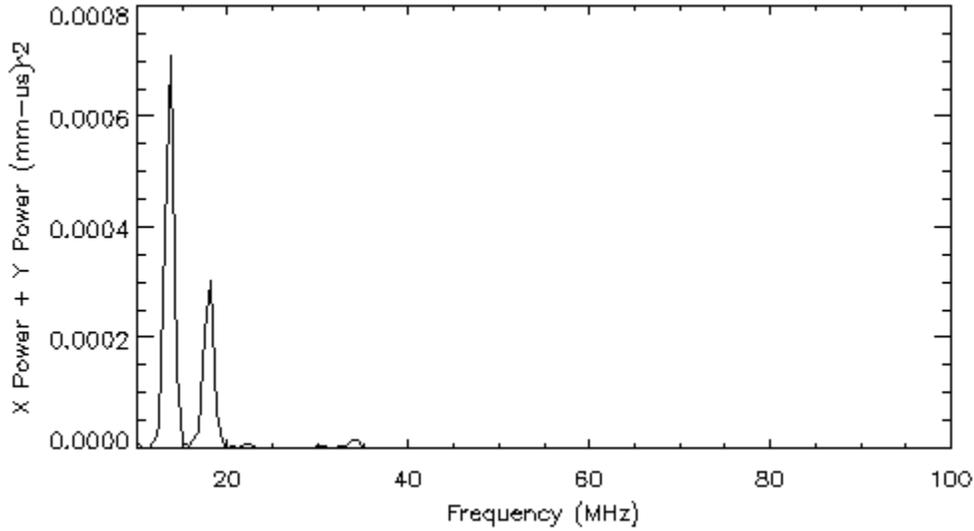

Figure 22. Power spectrum of motion shown in Figure 21 from 10 MHz to 100 MHz.

The highest frequency motion expected at the accelerator exit is BBU. The tunes for commissioning were designed to have strong enough fields to suppress the BBU to low amplitudes [10]. The transverse impedance of the refurbished cells measured at LBNL [11] is compared in Figure 23 with the transverse impedance of the legacy cells which were used for the BBU measurements in Ref. [10] .The expected saturated BBU gain for the relaxed tune as calculated from XTR simulation is shown in Figure 24.

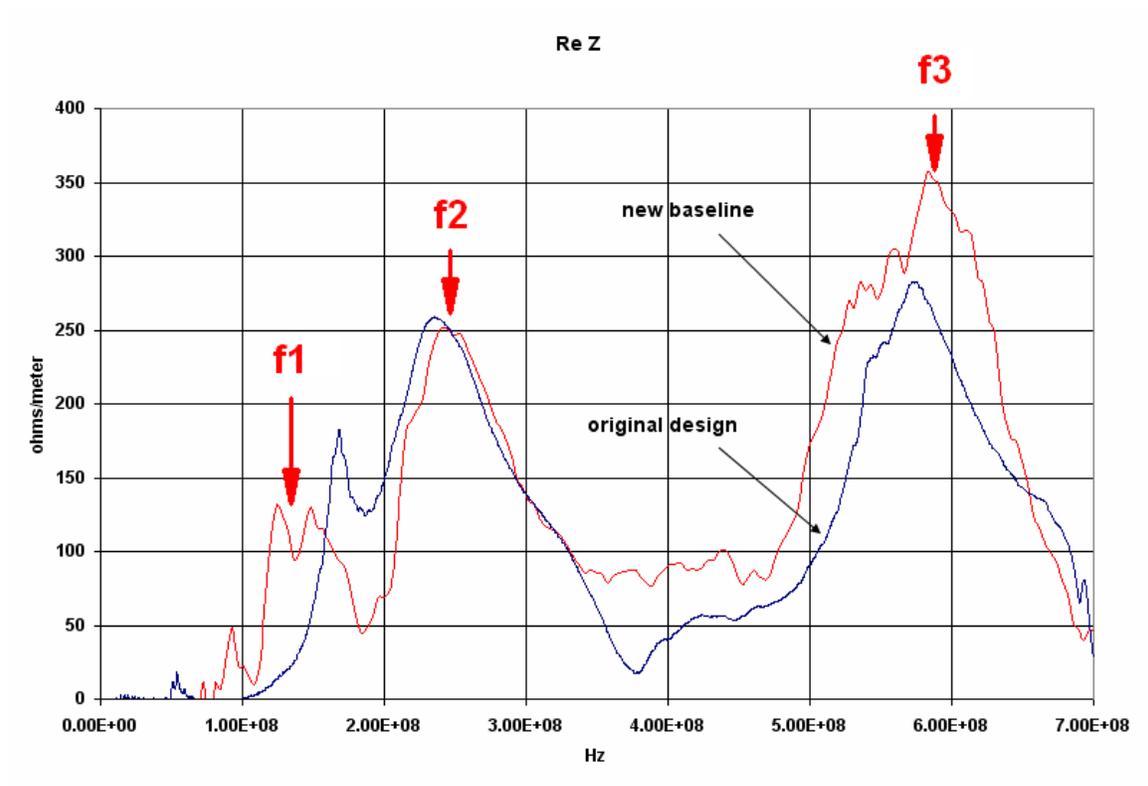

Figure 23. Transverse impedance of legacy and refurbished cells measured at LBNL [11].

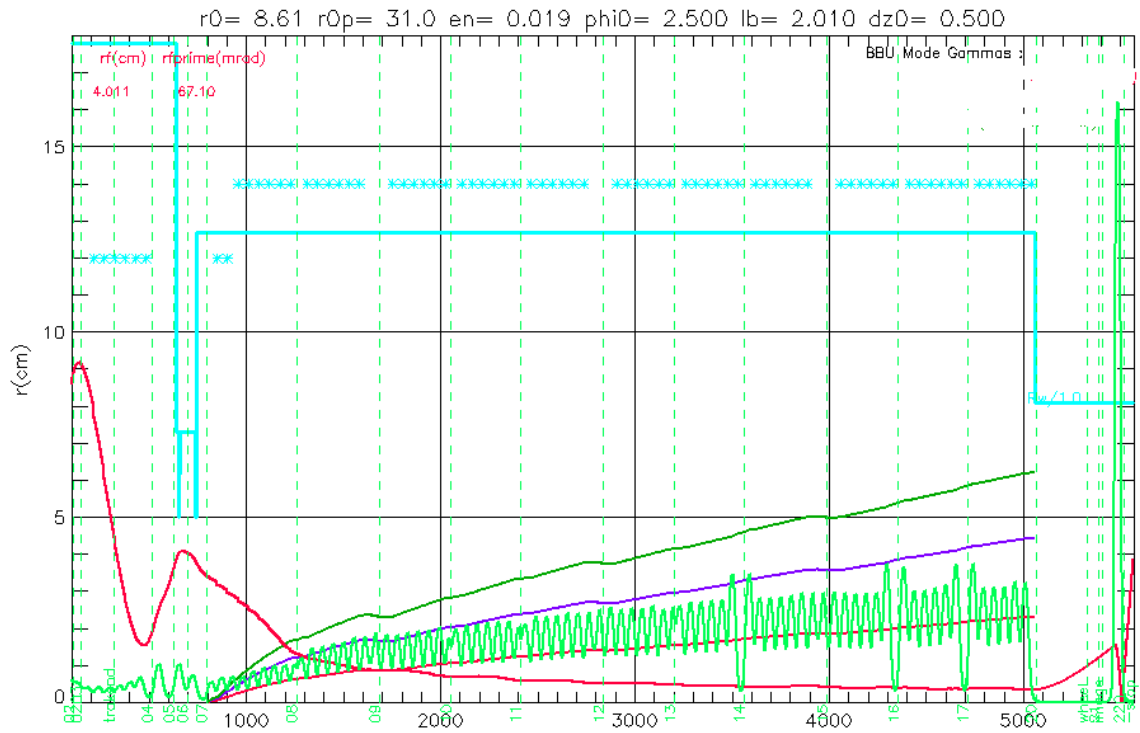

Figure 24. XTR envelope simulation of relaxed tune showing saturated BBU growth for the three dominant modes.

We recorded the beam position data at the accelerator exit (BPM20) at a rate of 5 Gs/s in order to have enough bandwidth to resolve even the highest frequency BBU mode that was measured at LBNL. Figure 25 shows the BBU amplitude at BPM20 during a 200-ns window near the end of the beam pulse, as was done in Ref. [10]. The peak-to-peak amplitudes of the BBU shown in Figure 25 (30-60 microns) appear to be slightly less than those shown in Figure 9 of Ref. [10] (40-100 microns), as predicted by comparing the saturated growth for both tunes.

Figure 26 shows a spectral analysis of the beam motion at the accelerator exit. In addition to the lower BBU frequency activity, the BBU activity at the highest frequency mode (~600 MHz) can be clearly seen. Figure 27 is a spectral analysis of the beam transverse velocity, which helps to enhance the higher frequency sensitivity. That the amplification at 600 MHz does not appear to be as strong compared with the lower frequency amplification as might be expected from the transverse impedance measurements is most likely due to the bandwidth limitations of our BPMs and cable recording system.

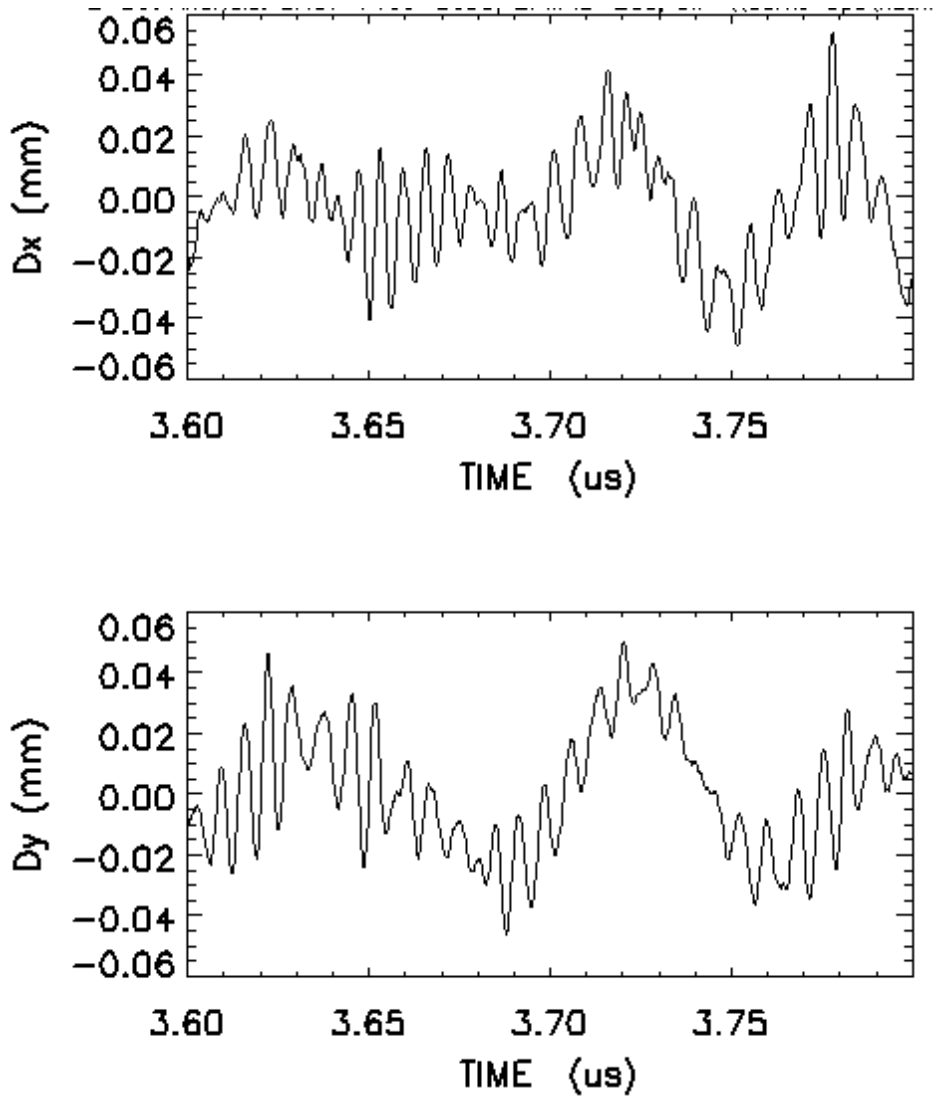

Figure 25. Beam motion at accelerator exit (BPM20) during a 200-ns window near the end of the 1.6-µs flattop.

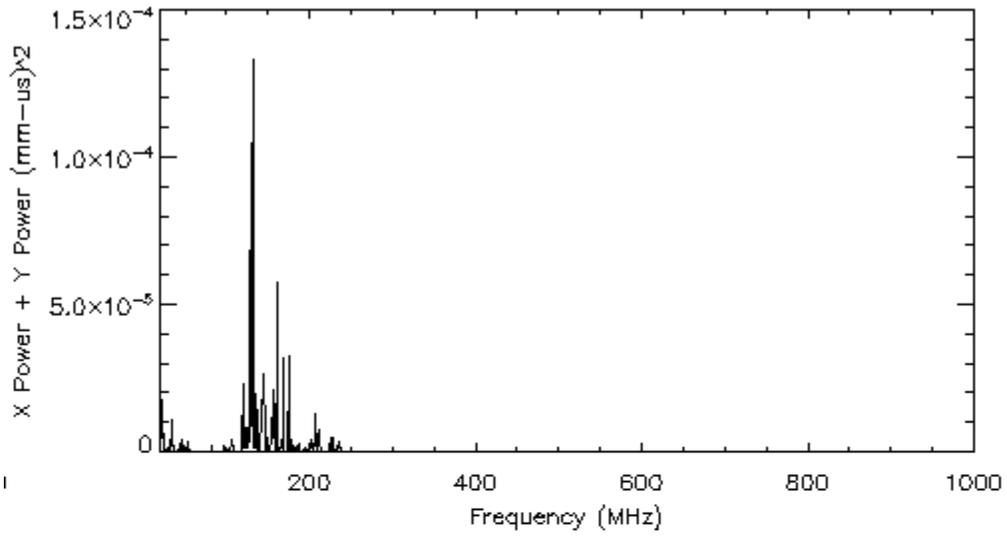

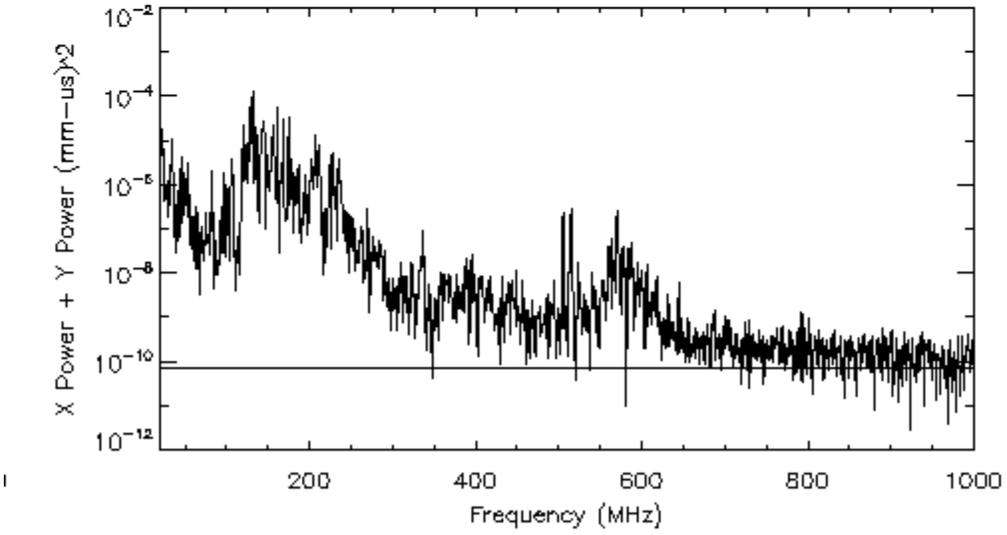

Figure 26. Spectrum of beam motion at BPM20 over the full 1.6-μs window shown in Figure 13 for shot 7169. Top: Plotted on linear scale. Bottom: Plotted on semi-logarithmic scale.

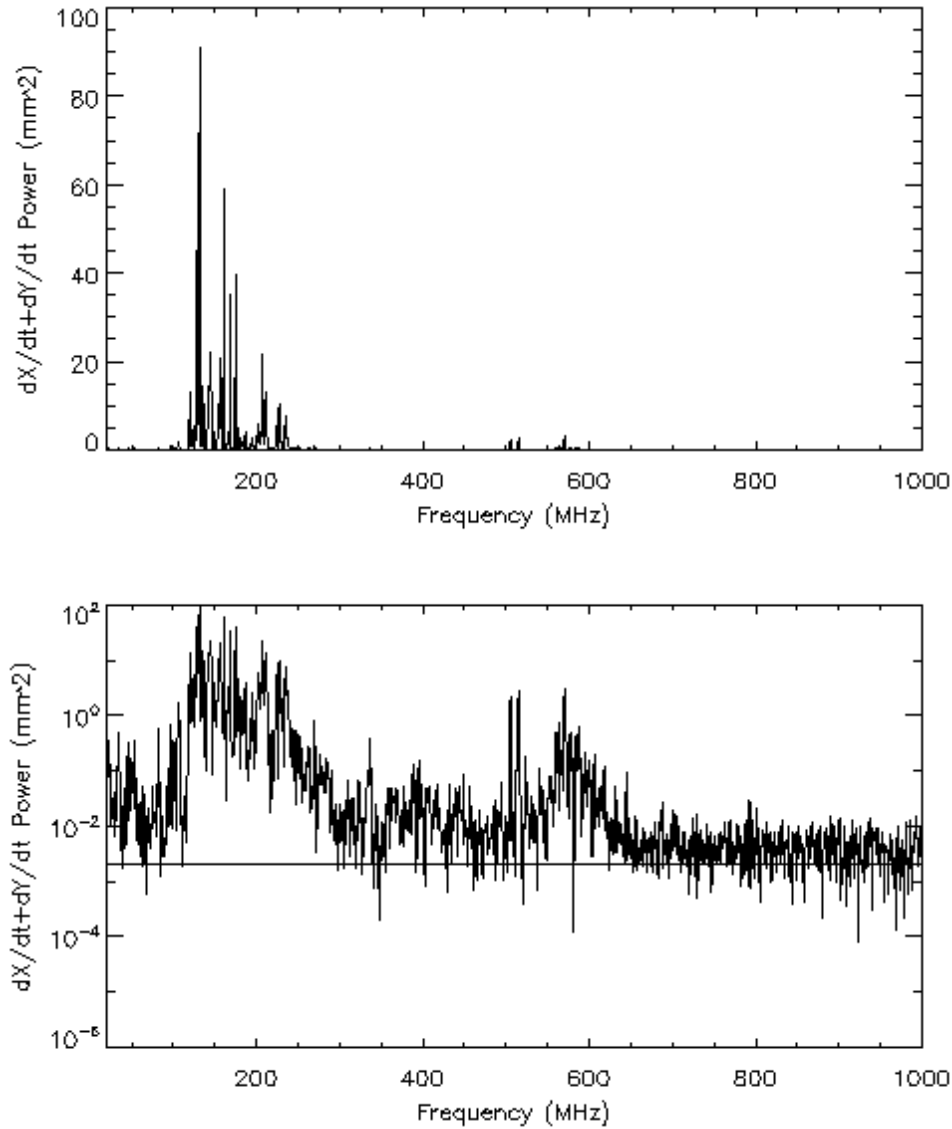

Figure 27. Spectrum of beam transverse velocity at BPM20 over the full 1.6-μs window shown in Figure 13 for shot 7169. Top: Plotted on linear scale. Bottom: Plotted on semi-logarithmic scale.